\documentclass[times, 12pt]{article}
\usepackage{setspace}
\usepackage{multirow}
\usepackage{rotating}
\usepackage{booktabs}
\usepackage{tikz, pgflibraryshapes}
\usepackage{epstopdf}

\def\corraddr#1{%
  \gdef\@corraddr{%
    \footnotetext[1]{\itshape Correspondence to: #1}\stepcounter{footnote}}}

\def\address#1{%
  \gdef\@address{%
    \footnotetext[0]{\noindent\rule{\textwidth}{5.0pt}\vskip 4pt\noindent\itshape #1}}}

\def\keywords#1{%
  \gdef\@keywords{{%
    \textbf{Keywords:}}\hspace{0.75em}\parbox[t]{32pc}{#1}}}

\begin{document}

\title{Prediction meets causal inference: the role of treatment in clinical prediction models}

\author{N van Geloven$^{1}$\footnote{Correspondence: PO box 9600, zone S5-P, 2300 RC Leiden, The Netherlands. E-mail: n.van\_geloven@lumc.nl}, SA Swanson$^{2}$, CL Ramspek$^3$, K Luijken$^3$,\\ M van Diepen$^3$, TP Morris$^{4}$, RHH Groenwold$^{1,3}$,\\ HC van Houwelingen$^{1}$, H Putter$^{1}$, S Le Cessie$^{1,3}$}

\footnotetext[1]{Department of Biomedical Data Sciences, Leiden University Medical Center, Leiden, the Netherlands}
\footnotetext[2]{Department of Epidemiology, Erasmus MC, the Netherlands and Department of Epidemiology, Harvard T. H. Chan School of Public Health, USA}
\footnotetext[3]{Department of Clinical Epidemiology, Leiden University Medical Center, Leiden, the Netherlands}
\footnotetext[4]{MRC Clinical Trials Unit, UCL London, United Kingdom}


\maketitle

\section*{Abstract}

In this paper we study approaches for dealing with treatment when developing a clinical prediction model. Analogous to the estimand framework recently proposed by the European Medicines Agency for clinical trials, we propose a `predictimand' framework of different questions that may be of interest when predicting risk in relation to treatment started after baseline. We provide a formal definition of the estimands matching these questions, give examples of settings in which each is useful and discuss appropriate estimators including their assumptions. We illustrate the impact of the predictimand choice in a dataset of patients with end-stage kidney disease. We argue that clearly defining the estimand is equally important in prediction research as in causal inference.


\section{Introduction}
Clinical prediction models provide individualized prognostic information that can be used to counsel patients about the likely course of their disease. Prediction models can also support treatment decisions. For instance, if a patient's risk of a poor outcome in the next year is relatively low, then she or he may not need treatment. If the risk of a poor outcome is high, additional preventive or curative treatments should be considered \cite{Hemingway2013}. But what is exactly meant by ``a patient's prognosis'' or ``a patient's risk'' here? Do we mean the risk assuming that no treatment is given, the risk under current standard treatment, the risk of experiencing the event in the time period before being treated, or something else? An increasing number of clinical prediction models are becoming available through websites and apps. However, for many of these models it is unclear how the risk they aim to capture relates to treatment.

The data sources used for the development of clinical prediction models often contain data on a mix of patients who did and did not receive treatments that affect the risk of the event of interest. This holds both for observationally collected data and for trial data. Completely untreated cohorts can sometimes be found in historical data collections, but these cohorts are rare and may not be relevant to current practice in other important ways. Using data solely from untreated patients can also lead to highly selected cohorts that are not generalizable. 
Dealing with treated patients is thus a challenge that needs to be addressed when developing a clinical prediction model. Baseline treatments can be included as predictors when developing, or externally validating, a prediction model, meaning that for a new patient the model can be used to predict the outcome, provided information is available about baseline treatment status and other predictors \cite{Groenwold2016}. However, for treatments that are initiated after baseline, there is no easy solution, nor is there a single question of interest \cite{Sperrin}. We do not know at baseline what treatments will be initiated later on and we cannot use future values as predictors. Moreover, the way treatments are dealt with during model development will have impact on how predictions can be used in future patients. Currently, ad hoc approaches are used where treatment initiation may be ignored, patients are censored at the moment they start or switch treatment, or even excluded completely from the development cohort. Such analysis choices are often reported as mere technical analysis issues \cite{Pajouheshnia2017}, but they may have a major impact on interpretation. The predictions resulting from such approaches might end up targeting a different risk than researchers intended. For example, censoring patients when they receive a transplantation when estimating survival among patients listed for liver transplantation has shown to overestimate the actual observed waiting list mortality \cite{Kim2006, Staplin2015}. The TRIPOD reporting guideline on development and validation of prediction models offers hardly advice on the issue. The only related point on the TRIPOD checklist is that one should describe the treatments that patients received, if relevant (item 5c) \cite{TRIPOD}.

The European Medicines Agency (EMA) has recently released a new guideline that provides a framework to deal with additional treatments started after baseline and other so-called intercurrent (post-baseline but pre-outcome) events in the context of clinical trials: the addendum to the ICH E9 guideline Statistical Principles for Clinical Trials \cite{EMA}. The guideline distinguishes five possible strategies: a `treatment  policy' strategy that follows the intention-to-treat principle and ignores any change in treatment after baseline, a `composite' strategy that includes the start of an additional treatment as part of the outcome definition, a `hypothetical' strategy that aims to assess the outcome in a scenario where no additional treatment would be given, a `principal stratum' strategy where the outcome is considered in a certain subset of patients who would never start the additional treatment independent of their allocated treatment, and a `while on treatment' strategy where the outcome is assessed in the time period up until the additional treatment is started. Together with the definition of the population, the outcome of interest and the effect measure, each strategy defines an estimand which is the target quantity that the trialists aim to estimate. Each estimand represents a different causal question. In the trial context, these questions relate to treatment effects. In prediction models, the aim is to assess patients' expected outcomes conditional on certain patient characteristics measured at baseline. We argue that, just like in trials, the quantity that a clinical prediction model targets should be unequivocally defined. Therefore we map the E9 estimand framework to the context of prediction models, proposing a predictimand framework.

Our focus in this manuscript is on prognosis over time, so on time-to-event or failure-time outcomes. We consider treatments that are initiated during follow-up. Previous studies on estimands in prediction research considered point (time-invariant) treatments \cite{Groenwold2016} and binary outcomes \cite{Sperrin}. Recently, Pajouheshnia and colleagues discussed analysis methods for time-to-event prediction of untreated risk in the presence of time-dependent treatment \cite{Pajouheshnia}. Here we will extend that work by considering different questions that can be assessed by prediction models and that will be of interest in different applied settings. We formulate these strategies analogous to the trial estimand framework from the E9 addendum. We provide a formal definition of the prediction estimands matching these questions. For each of the estimands, we discuss key assumptions for estimation and list common estimators. We illustrate the impact of the predictimand choice in a dataset of patients with end-stage kidney disease.

\section{Notation}
To simplify, we consider only one treatment ($A$) that is related to the event of interest. Patients are all event-free and without this treatment at time zero. At that moment we collect baseline covariates $X(0)$, which will be used to determine the prognosis of the patient. $T$ is the time to the event of interest. Some patients will start treatment $A$ over time, with $V$ the time to treatment start and $A(t)$ the time dependent treatment indicator. In principle, $A(t)$ could switch between 0 (no treatment) and 1 (treatment) multiple times over the follow up, but, to enhance readability, in the following we will assume that once patients initiated treatment, they stay in the treated condition throughout. For patients who experience the event of interest before treatment start, $V$ is latent. Both $T$ and $V$ can be censored by end of study or loss to follow up which we assume for simplicity to be non-informative censoring mechanisms. In some studies, patients are no longer followed for the event of interest after treatment initiation, for instance in a registry of patients on dialysis that no longer follows patients after a kidney transplantation. This situation is depicted in Figure \ref{fig:stattrbl}a. If follow up on the event of interest does continue after a new treatment is initiated, we are in the setting depicted in Figure \ref{fig:stattrbl}b. The two figures are not causal directed acyclic graphs, but could be viewed as state transition diagrams of a multistate model \cite{Putter}. Figure \ref{fig:stattrbl}a is similar to a competing risks setting and Figure \ref{fig:stattrbl}b is similar to that of an illness death model, with the intermediate state in our case not disease but treatment. The pace at which patients initiate treatment is denoted by the transition intensity $\alpha(t,H_t)$, where $H_t$ is the history of the patient up to time point $t$. Later on, we will distinguish between situations where the treatment decisions are only based on patients'  baseline prognostic covariates, i.e., $H_t=\{X(0)\}$ and situations where the treatment decisions are also based on prognostic markers that evolve over time, i.e., $H_t=\{X(s);s\leq t\}$.


\begin{figure}
a
\begin{center}
    \begin{tikzpicture}[>=stealth]
    \begin{scope}[shape=rectangle,fill=blue!20]
        \tikzstyle{every node}=[draw,fill,text width=3cm,text badly centered]
        \node (Alive) at (0,0) {Event-free};
        \node (One) at (8,0) {Event of interest};
        \node (K) at (4,3) {Treatment $A$};
    \end{scope}
    \draw [->, shorten >= 2pt, shorten <= 2pt, solid] (Alive) --node[above, sloped]{$\alpha(t,H_t)$} (K);
    \draw [->, shorten >= 2pt, shorten <= 2pt, solid] (Alive) --node[above]{} (One);
    \end{tikzpicture}
    \end{center}
    b
    \begin{center}
    \begin{tikzpicture}[>=stealth]
    \begin{scope}[shape=rectangle,fill=blue!20]
        \tikzstyle{every node}=[draw,fill,text width=3cm,text badly centered]
        \node (Alive) at (0,0) {Event-free};
        \node (One) at (8,0) {Event of interest};
        \node (K) at (4,3) {Treatment $A$};
    \end{scope}
    \draw [->,  shorten >= 2pt, shorten <= 2pt, solid] (Alive) --node[above, sloped]{$\alpha(t,H_t)$} (K);
    \draw [->, shorten >= 2pt, shorten <= 2pt, solid] (Alive) --node[above]{} (One);
    \draw [->,  shorten >= 2pt, shorten <= 2pt, solid] (K) --node[above]{} (One);
    \end{tikzpicture}
    \end{center}
    \caption{Graphical representation of the studied situation. Follow up on the event of interest may stop (panel a) or continue (panel b) after treatment initiation.}
    \label{fig:stattrbl}
    \end{figure}

\section{Predictimands}

Below we describe four strategies for how to deal with treatment initiation after baseline in the development of a prediction model. We formulate the interpretation of the resulting prediction estimands, give examples of settings where they are useful and discuss how they apply to new patients that were not used for development of the predictions, i.e., their generalizability \cite{Justice}. The four strategies described have analogous estimands in the E9 addendum. The fifth -principle stratum- estimand described in the addendum was not mapped to the prediction setting as it refers to a counterfactual subpopulation that has no immediate analogue to the prediction setting. In Section \ref{sec:estimators} we focus on assumptions needed for estimation of the predictimands and list some common estimators. An overview is presented in Table~\ref{tab:predictimands}.

\begin{sidewaystable}
\centering
\caption{Overview of the four strategies of dealing with treatment initation after baseline. $T$ time to event of interest; $V$ time to start of treatment; PCI percutaneous coronary intervention; CABG coronary artery bypass grafting; IVF in vitro fertilization.}
\label{tab:predictimands}
\begin{tabular}{p{0.16\linewidth}p{0.19\linewidth}p{0.23\linewidth}p{0.23\linewidth}p{0.18\linewidth}}
\\
Strategy & Estimand &  Example & Estimators & Key assumptions \\
\hline
ignore treatment   & risk of the event, regardless of treatment & risk of cardiovascular events where some patients will initiate statins according to routine-care prescriptions& survival model for $T$, do not censor at $V$ & treatment assignment policy in application setting similar to development data \\
\\
composite          & risk of the event or treatment initiation & risk of a composite of cardiovascular death, myocardial infarction and treatment with revascularisation (PCI or CABG)     & survival model for min($T,V$)   &   treatment assignment policy in application setting similar to development data \\
\\
while untreated     & risk of the event occurring before treatment is started   & risk of dying while on the waiting list for a liver transplant & competing risks methods & treatment assignment policy in application setting similar to development data \\
\\
hypothetical       & risk of the event if treatment were never started  & risk of a natural pregnancy without IVF treatment & survival model for $T$, censor at $V$ or include treatment as time-dependent covariate in the model and set to 0 when predicting  & exchangeability, consistency and positivity  \\
\hline
\end{tabular}
\end{sidewaystable}

\subsection{``Ignore treatment'' strategy}
In this strategy, treatment initiation is considered part of standard practice. The value for $T$, the time to event of interest, is used regardless of whether patients start treatment or not. So $V$, the time to treatment, is not used in any way. This is analogous to the ``treatment policy'' strategy in the E9 guideline \cite{EMA}. It has also been described as ``simply ignore treatment'' \cite{Groenwold2016}. The risk that is estimated equals:
\begin{equation}
\label{eqn:ignore}
P(T\leq t_{\rm{hor}} | X(0)),
\end{equation}
i.e., the risk of the event of interest occurring before a time horizon $t_{\rm{hor}}$ under the treatment practice inherent to the development dataset. An example of where this strategy was used, is in the development of QRISK3, a risk prediction algorithm that targets a person's risk of a heart attack or stroke over the next 10 years \cite{QRISK3}. The algorithm was developed using individuals who did not use statins at baseline, but statin use after baseline was ignored, as discussed in \cite{Sperrin}. The calculated risks therefore belong to a population where some individuals will receive statins during follow up. The algorithm will only be generalizable to new patient groups if in those groups the same treatment assignment policy is used as in the development cohort. This implies that for all subgroups defined by the predictors in the prediction model, a similar proportion of patients should initiate treatment as in the development set. Provided the treatment is effective in reducing the risk of the outcome, the risk calculated with the ``ignore treatment'' strategy will be lower than the untreated risk, that is, the risk for a patient who will not be treated during follow-up. If the ignore treatment predictimand is falsely interpreted as untreated risk and used for future decisions on prescribing (statins), it will underestimate the true untreated risk and this could lead to undertreatment in new patients \cite{Peek}. The ``ignore treatment'' strategy requires continued follow up after treatment initiation, so it cannot be used in the study design depicted in Figure \ref{fig:stattrbl}a.

\subsection{``Composite'' strategy}
In the second strategy, treatment is combined into a composite outcome together with the event of interest. In such a ``composite'' strategy we target
\begin{equation}
\label{eqn:composite}
P(\min(T,V)\leq t_{\rm{hor}}|X(0)),
\end{equation}
i.e., the risk of the event of interest or the treatment occurring before time $t_{\rm{hor}}$. In this strategy the treatment is integrated in the clinical outcome. An example 
is predicting the composite of cardiovascular death, myocardial infarction and treatment with revascularisation (PCI or surgery) \cite{Hicks}. Including the treatment in the outcome may seem a somewhat artificial way of dealing with treatment and may lead to a less well interpretable outcome, but some use cases exist. A ``composite'' strategy can for instance be used when the treatment has very likely prevented an imminent occurrence of the event of interest and treatment can be viewed as a proxy of the event (i.e., without PCI or surgery a patient would develop a myocardial infarction). Other settings where a composite outcome can be useful is when a poor outcome is more clearly captured by its consequence of needing treatment than by giving a precise description of poor health status. For example, Von Dadelszen et al. used a composite outcome including amongst others receiving infusion of a third parental hypertensive drug, intubation, transfusion with any blood product and dialysis, when predicting severe maternal outcomes in pre-eclampsia \cite{Dadelszen}. A third use case for the composite strategy is predicting the chances of a good outcome (one minus composite) defined as staying event-free without requiring additional treatment. The ``composite'' strategy does not need continued follow up after treatment initiation. Applying a composite prediction estimand to new patients requires similar treatment assignment policies as in the development cohort. For instance, if in the new setting where the predictions are applied, patients in a certain subgroup are treated more often than similar patients in the development cohort, then the predictions of the combined outcome will be lower than the true probabilities in that subgroup (miscalibration).

\subsection{``While untreated'' strategy}
In the ``while untreated'' strategy, we are only interested in the event of interest if it happens before treatment is started. Events occurring after treatment start do not count as events. We target the risk of the event of interest occurring before time $t_{hor}$ and before treatment is started.
\begin{equation}
\label{eqn:cuminc}
P(T\leq t_{\rm{hor}}, T<V|X(0)).
\end{equation}
This prediction estimand is well-known from competing risks analysis. It is often referred to as cumulative incidence.  Starting treatment is then considered a competing event that precludes observing the untreated event of interest. Cumulative incidence has also been referred to as the absolute risk, actual risk, crude probability, crude cumulative incidence function, absolute cause-specific risk or subdistribution function \cite{Grunkemeier, Geskus, Pfeiffer}. It has recently been conceptualized as the ``risk without elimination of competing events'' \cite{Young}. The analogous name for this strategy in the E9 addendum is ``while on treatment'', referring to the response of the patient during the period where patients are still on their originally assigned treatment. Note that what distinguishes this strategy from the ``ignore treatment'' strategy is that at the moment treatment is started, the event of interest will by definition not occur anymore. An example is estimating the risk of dying while on the waiting list for a liver transplant \cite{Kim2006}. Typically this strategy will be of interest if the treatment is not freely available, but limited due to waiting lists or other logistical constraints. Also for this strategy, predictions on new patients are only well calibrated if the assignment policy of treatment is similar to the development cohort. If in the application setting a subgroup is treated more often or sooner than in the development set, the predictions will overestimate the true ``while untreated'' risk in this subgroup.

\subsection{``Hypothetical'' strategy}
In this strategy we envision a world where treatment does not exist. We aim to estimate the untreated risk before time $t_{\rm{hor}}$:
\begin{equation}
\label{eqn:hyp}
P(T^{v = \infty} \leq t_{\rm{hor}} | X(0)),
\end{equation}
where $T^{v=\infty}$ represents the counterfactual time to the event of interest if $V$ is set to infinity, i.e., in a hypothetical world where treatment $A$ is eliminated. Like the ``while untreated'' strategy, this strategy too has an analogue in the competing risk literature. Young et al refer to this risk as the ``risk under elimination of competing events'' \cite{Young}. It has been referred to as the marginal cumulative incidence, net risk, or pure risk \cite{Geskus, Pfeiffer}. The risk quantifies how likely the event of interest would be if nobody were to receive treatment. Since we are only interested in the risk up to $t_{\rm{hor}}$, we could similarly have used $T^{v>t_{\rm{hor}}}$ instead of $T^{v=\infty}$. In the E9 addendum this strategy is similarly referred to as the ``hypothetical'' strategy.

An example application is estimation of the ‘risk’ of a natural pregnancy without use of assisted reproductive techniques such as IVF \cite{VanGeloven}. Other hypothetical scenarios (e.g., what if treatment is started after one year) could in principle also be targeted. In fact, if we could set $v$ exactly according to the function that clinicians used in the development cohort to determine when to start treatment, this estimand would reduce to the ``ignore treatment'' estimand in (1). But here we only discuss the hypothetical untreated risk further. Estimating the hypothetical untreated risk is challenging (see Section \ref{sect:estimators_hyp}), but, once constructed, it is readily generalisable to new patients when posing the question what would happen if the new patient is never treated.  This untreated/baseline risk is useful to inform decisions on treatment $A$.  Note that the three other strategies (ignore treatment, composite and while untreated) cannot be used to inform the decision to start treatment $A$, as this might lead to something that has been described as the `prediction paradox': predictions influencing behaviours (i.e., treatment decisions) that in turn invalidate predictions \cite{Peek}. These three predictimands will be miscalibrated if the treatment decisions made in new patients differ from those in the development cohort.

\section{Estimators and their assumptions}\label{sec:estimators}

In this section we focus on key assumptions and design elements that are necessary for estimating each predictimand. Without being exhaustive, we link the estimands to common estimators.

\subsection{``Ignore treatment'' strategy} Since in this strategy starting treatment after baseline is ignored, standard time to event regression methods may be used to relate the event of interest to the covariates $X(0)$. For instance, one could use Cox regression models combined with the nonparametric Breslow (or Efron) estimator for the baseline hazard or flexible parametric survival models. The main assumption here is that of non-informative censoring for reasons like loss to follow up or end of study. Particular methods may additionally assume proportional hazards for the covariates. Note that since we do not censor at the moment of treatment start, this strategy requires continued follow up after treatment initiation.

\subsection{``Composite'' strategy} Also with this strategy the analysis is relatively straightforward and can be done with any chosen survival regression technique. Either occurrence of the event of interest or the occurrence of treatment counts as an event, whichever comes first. When one would use a Cox-like model, the assumption of proportionality of covariate effects should hold for the composite outcome, which may be less likely than for single outcomes. Also, the non-informative censoring by loss to follow up or end of study should hold in relation to the composite outcome. Continued follow up after treatment initiation is not needed.

\subsection{``While untreated'' strategy} Estimation of cumulative incidence as expressed in (\ref{eqn:cuminc}) can be done with competing risks methods. Without covariates and without censoring for other reasons like loss to follow up or end of study, cumulative incidence is estimable by the number of patients with the event of interest divided by the total number of patients at baseline. With covariates and censoring for reasons like loss to follow up or end of study, the estimation can be done in various ways, see for instance \cite{Putter, FineGray, Scheike, Nicolaie, Sachs}. Continued follow up after treatment initiation is not needed.

\subsection{``Hypothetical'' strategy}\label{sect:estimators_hyp} Estimation of (\ref{eqn:hyp}) is challenging and relies on strong assumptions regarding the treatment assignment policy in the development data. These assumptions are similar to those that are needed for identifying the causal treatment effect of $A$, which makes sense since the strategy implicitly imposes a counterfactual or potential outcome version of $A$ into the estimand of interest. Three key assumptions are required: exchangeability, consistency and positivity \cite{HernanRobins}. The first one, exchangeability, is often the most challenging. It is sometimes called the `no unmeasured confounding' assumption and requires that we have measured and appropriately corrected for (a sufficient subset of) the variables that both influenced the treatment decisions and are prognostic for the event of interest \cite{HernanRobins}. This typically requires measuring time-dependent covariates since updated measurements of risk factors are very likely to influence treatment decisions. The second assumption, consistency, is often described as observed outcomes being equal to counterfactual outcomes. It means that in the hypothetical world where treatment is eliminated, a patient's untreated risk is the same as her or his untreated risk in the real world. If knowledge of the unavailability of treatment changes the risk behaviour of patients, this assumption does not hold. This second assumption is typically not prohibitive; in some causal frameworks such changing risk behaviour is taken as a definition of the patient population \cite{Pearl}. The third assumption, positivity, means that we have observed a non-zero number of treated and untreated patients in our data for all covariate patterns during the time horizon that we want to use in our predictions (or have observed a sufficient number so as to smooth over gaps with modelling assumptions). For instance if all patients with a certain characteristic are treated after 1 year, then we don't have information for estimating untreated outcomes beyond one year for such patients. It may well be that a shorter prediction horizon $t_{\rm{hor}}$ has to be chosen to fulfil the positivity assumption.

 The analysis approach for the ``hypothetical'' strategy depends on how treatment decisions were made for the patients in the development data and on whether or not post-treatment follow up is used. Below, we sketch four analysis approaches. We first discuss two settings where it is assumed that in the development dataset treatment decisions were only based on baseline covariates $X(0)$, i.e., prognostic patient characteristics that are known at the moment we want to make the prediction. In most healthcare settings it is quite implausible that risk factor progression after baseline doesn't influence treatment decisions, but we include these options to indicate the limitations of common estimation approaches. Then, we discuss two settings where risk factor progression is accounted for. There are two analysis approaches to target the hypothetical risk: we may stop follow up when treatment is started, by censoring the time to event of interest at the moment treatment starts (Figure \ref{fig:stattrbl}a), or follow up data after the start of treatment may be included (Figure \ref{fig:stattrbl}b). As noted before, in some studies no follow up information is collected after treatment initiation, in which case only the censoring option remains. In the censoring approach positivity is only needed for untreated individuals, i.e., if for some covariate patterns no patients are treated, this is not per se a problem for estimating the untreated risk.

Several other estimation approaches than the ones described below have been proposed for the hypothetical untreated risk, each with their own assumptions. For instance using g-formula \cite{Young}, copulas \cite{Zheng, Escarela} or multiple imputation \cite{Hsu, Jackson}. We refer to the mentioned references for further details on these methods.

\subsubsection{Baseline covariates, censoring}
 In the situation sketched in Figure \ref{fig:stattrbl}a, time to event is censored at treatment start. The main assumption of this approach is that censoring by treatment is non-informative, conditional on the baseline prognostic covariates in the prediction model ($X(0)$). In other words, censoring the follow up at treatment start only gives a valid estimate of the hypothetical untreated risk if baseline prognostic factors that relate to treatment are included in the prediction model and treatment start is independent of changing values of prognostic markers during follow-up, i.e., $H_t=\{X(0)\}$. This is a strong and often implausible assumption that cannot be tested on the data. Such an assumption can only be verified with those who were responsible for the treatment decisions.


\subsubsection{Baseline covariates, modelling}\label{subsec:baseline} When follow up after treatment start is available, this follow-up information can be used in the development of the prediction model. Treatment $A$ can be added as an additional, time dependent, covariate to the prediction model. Then a prediction under the ``hypothetical'' strategy of no treatment can be obtained by setting $A(t)=0$ for all $t$ in the prediction horizon ($t\leq t_{\rm{hor}}$). This approach is valid under the same strong and often implausible assumption as the first approach that we described. Only in case treatment choices were solely based on collected prognostic baseline factors that are included in the prediction, the model including $X(0)$ is sufficient to get an unconfounded estimate of the effect of $A$. There is a caveat to this approach of modelling the treatment effect. The advantage of using a longer period of follow up is at the expense of having to model the effect of the treatment on the event of interest. Therefore, the functional form of the treatment effect should be carefully chosen: the assumption of proportional hazards between the treated and untreated patients over time and the presence of possible interactions between patient characteristics and treatment should be checked. 

 \subsubsection{Time varying covariates, censoring} We now turn to the more realistic situation where time-varying prognostic patient characteristics have additionally influenced the treatment decisions, i.e., $H_t=\{X(s),s\leq t\}$. For example when the condition of patients has been monitored by repeatedly measuring their blood values and these measurements have influenced the decisions about treatment initiation. In the censoring case (Figure \ref{fig:stattrbl}a), inverse probability of censoring weighting (IPCW) could be used. This approach assumes that treatment start is independent of the future untreated risk conditional on baseline covariates $X(0)$ and time dependent covariates $X(t)$. $X(t)$ is used to estimate time-varying conditional probabilities of starting treatment. By assigning weights to patients that are inversely proportional to their conditional probability of not yet being treated, a weighted population is created that (under the assumptions of exchangeability, consistency and positivity) mirrors the pseudo-population that would have been observed in the absence of treatment \cite{RobinsFinkel, Cole,  Matsuyama,Howe, Young}. Applying inverse probability weighting when estimating a survival model for the event of interest with censoring at treatment start thus corrects for the informative censoring related to $X(t)$ .


\subsubsection{Time varying covariates, modelling} When post-treatment follow up is used to model the effect of treatment on the event of interest, adding both $X(t)$ and $A$ in the survival model similar to Section \ref{subsec:baseline}, will in the presence of time varying covariates not yield a useful prediction model, because for a new patient $X(t)$ will not be known when predicting at baseline. The hypothetical untreated risk to estimate from Figure \ref{fig:stattrbl}b is similar to what is called the `controlled direct effect' in mediation analysis, when setting treatment as the mediator at a fixed zero level (no treatment) \cite{Goetgeluk}. 
 A potential solution is to fit a marginal structural model with inverse probability of treatment weighting (IPTW) to break the link between $X(t)$ and $A(t)$ \cite{RobinsHernan}. Using these weights, a model containing $X(0)$, $A(t)$ and potential interactions as predictors can be fitted and one can estimate the hypothetical risk from this model setting $A(t) = 0$. Details on implementing this approach for prediction modelling in both logistic and time-to-event models can be found in \cite{Sperrin}. We again have to assume correct specification of the treatment effect.

\section{Data application}
In this section, we study the four proposed prediction estimands in data from the Netherlands Cooperative Study on the Adequacy of Dialysis (NECOSAD) \cite{Ramspek}. This study is a multicenter cohort study in which patients with end stage renal disease were included at dialysis initiation if they were 18 years or older and had no previous kidney transplantation and no previous dialysis. The NECOSAD study was approved by the local medical ethics committees and all patients gave informed consent. 
Patients were followed until renal transplantation, death or end of study. Here we consider death the event of interest and renal transplantation is the treatment that may be initiated at some point in time after baseline. As in NECOSAD patients were not followed after renal transplantation, information on death after transplantation was retrieved by linking the NECOSAD data to the Dutch registry of renal replacement therapy, RENINE (Registratie Nierfunctievervanging Nederland) \cite{Renine}. Patients were included between 1997 and 2007, and followed until February 1, 2015. Patients who were not coded as deceased or lost to follow up in RENINE were assumed to be alive at end of follow up. Our initial data set contained n=2051 patients. We removed 6 patients with missing information on age, yielding 2045 patients in our analysis. For 43 patients who were still alive at the last follow up visit of the NECOSAD study, no link to the registry could be made and we censored time to death and, where applicable, time to transplantation for these patients at their last follow up in NECOSAD. The median time in follow up of the 2045 patients was 5.1 years. In this period, 749 patients received a kidney transplant, 1470 patients died of whom 248 after transplantation. Age and baseline dialysis type (hemodialysis (HD) versus peritoneal dialysis (PD)) were used as baseline predictors of mortality ($X(0)$). With HD, blood is pumped out of the patient's body and filtered by an artificial kidney machine. With PD, cleansing fluid is pumped into the patient's abdominal cavity and the lining of the abdomen acts as a natural filter to wash out waste and toxins. As this example is used for illustration purpose, we did not include more baseline variables in the prediction model. Additionally, for estimating the hypothetical prediction estimand, we used the following time dependent covariates $X(t)$ as predictors of treatment: Charlson comorbidity score, BMI and calcium blood values, which were measured at 6 months intervals. We estimated the mortality risk over a time span of 10 years, given age (as a continuous variable) and baseline dialysis type. We used the packages \verb|survival|, \verb|mstate|, and \verb|ipw| of the R statistical software \cite{R}. Our analysis code along with a simulated dataset can be found in the Supplementary Materials. The different predictimands were estimated as follows:

\begin{itemize}
\item The ``ignore treatment'' strategy targets the total mortality risk, regardless of whether patients did or did not receive a transplantation.  For estimation, we used a Cox proportional hazards model with the non-parametric baseline hazard estimated using the approach proposed by Efron (further on referred to as Cox-Efron) \cite{Efron}. Death was defined as the event, age and dialysis type as baseline covariates and we censored the patients alive at the moment of last follow up. Note that for estimating the ``ignore treatment'' risk, follow up for death after transplantation is needed, which was retrieved by linking the NECOSAD data to the Dutch Renal Registry.
\item With the ``composite'' strategy, we estimate the risk of either dying or receiving a transplantation. To this end, transplantation and death were combined as composite event in a Cox-Efron analysis, again with age and dialysis type as baseline covariates and censoring those alive at the moment of last follow up. Studying a composite outcome in this situation can be informative, e.g., for policy makers, to know how long patients will likely stay alive and without transplantation and thus remain on dialysis treatment. For estimating the ``composite'' risk, follow up after transplantation is not needed.
\item The ``while untreated'' strategy assesses the risk of dying before receiving a transplantation. To estimate this risk, we fitted two cause-specific Cox-Efron models: one model with death as event, age and dialysis type as baseline covariates and censoring at time of transplant or at moment of last follow up alive, and one model with transplant as event, age and dialysis type as baseline covariates and censoring at death and at last follow up alive. The two cause specific hazard models were used to obtain the cumulative incidence for death \cite{Putter}. Follow up after transplantation is not needed for the ``while untreated'' risk.
\item In the ``hypothetical'' strategy we estimate the risk of dying if no transplantation is performed. We followed the four different estimation methods described in Section \ref{sect:estimators_hyp}.
\begin{itemize}
\item First, we fitted a Cox-Efron model for death, with age and dialysis type as baseline covariates and where event times were censored when the patient received a transplantation or at the end of follow up. This model was then used to predict the mortality risk over time. This approach assumes that the decisions on transplantation were based only on on age and dialysis type. Follow up after transplantation is not needed in this case.
\item Second, transplantation was included as a time-dependent covariate in a Cox-Efron model with age and dialysis type as baseline covariates, again assuming that only these two baseline covariates drove the transplantation decisions. To model the effect of transplantation correctly, we explored whether adding interactions between transplantation and baseline covariates improved the model, but it did not. Since the transplantation effect seemed to change over time, we used a time varying coefficient for treatment according to a step function (with jumps at 3 and 8 years, chosen by visual inspection of the Schoenfeld residual plot). The hypothetical untreated risk was then estimated by setting $A(t)=0$. For this second approach where the effect of transplantation is modelled, we needed the additional follow up of death after transplantation.
\item Third, we repeated the first analysis where we censored at treatment start, now applying inverse probability weighting to correct for time-dependent covariates that might have additionally influenced the transplantation decisions. Stabilized weights were estimated based on two Cox-Efron models with transplantation as event: a denominator model including Charlson comorbidity score, BMI and calcium blood values as covariates and a numerator model with only an intercept. Some missings occurred in the time dependent covariates and we performed a single imputation method for each using a linear mixed model with follow up time, age and dialysis type as fixed factors and a random intercept. For patients who did not have any measurement of these covariates (1 for Charlson score, 17 for BMI, 83 for calcium), we imputed the median of the other patients.
\item Fourth, we repeated the second analysis where we modelled the effect of transplantation, applying the same inverse probability weights as in the third approach.
\end{itemize}
\end{itemize}

\begin{figure}
\centering
\includegraphics[scale=0.45]{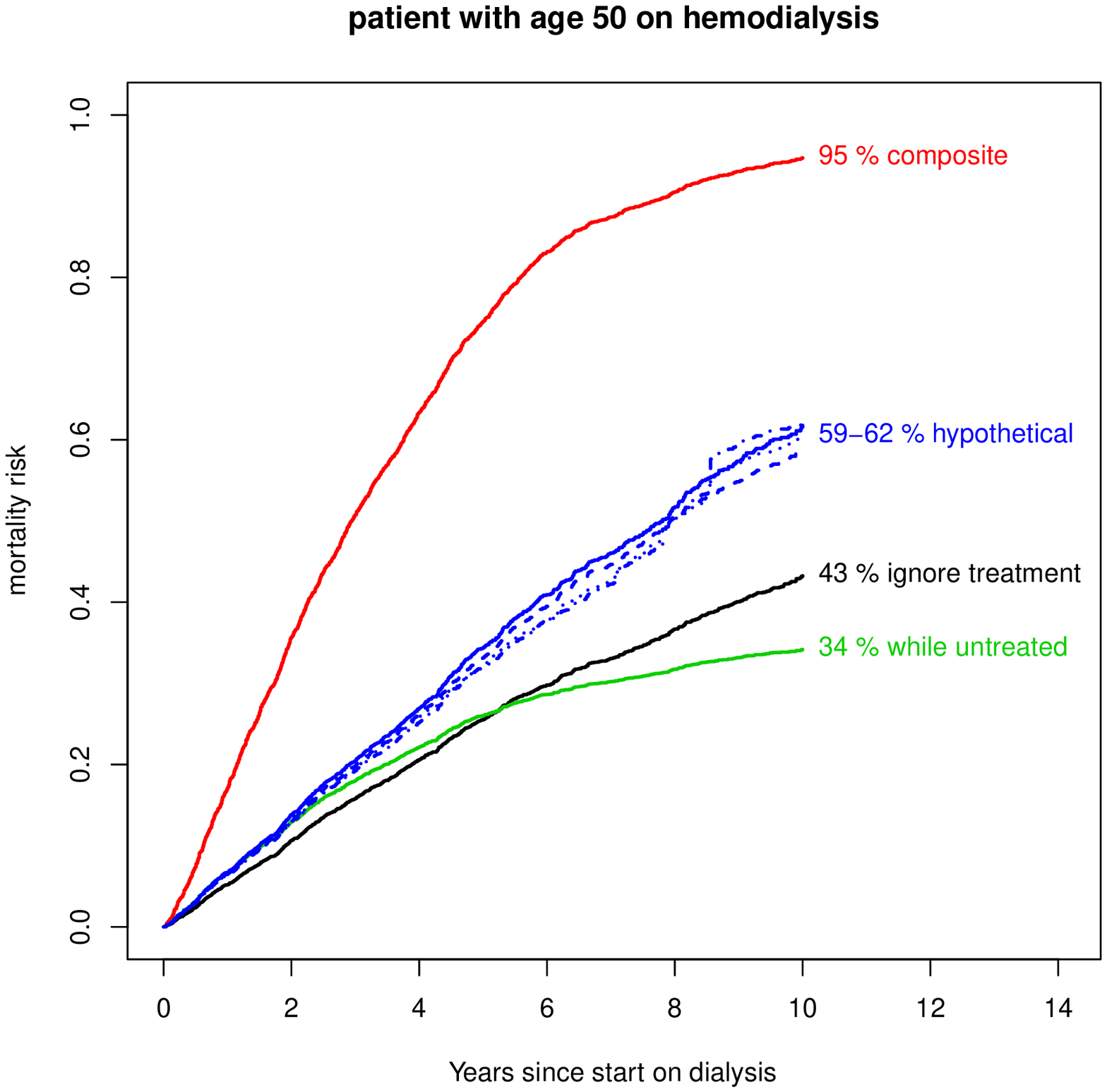}
\includegraphics[scale=0.45]{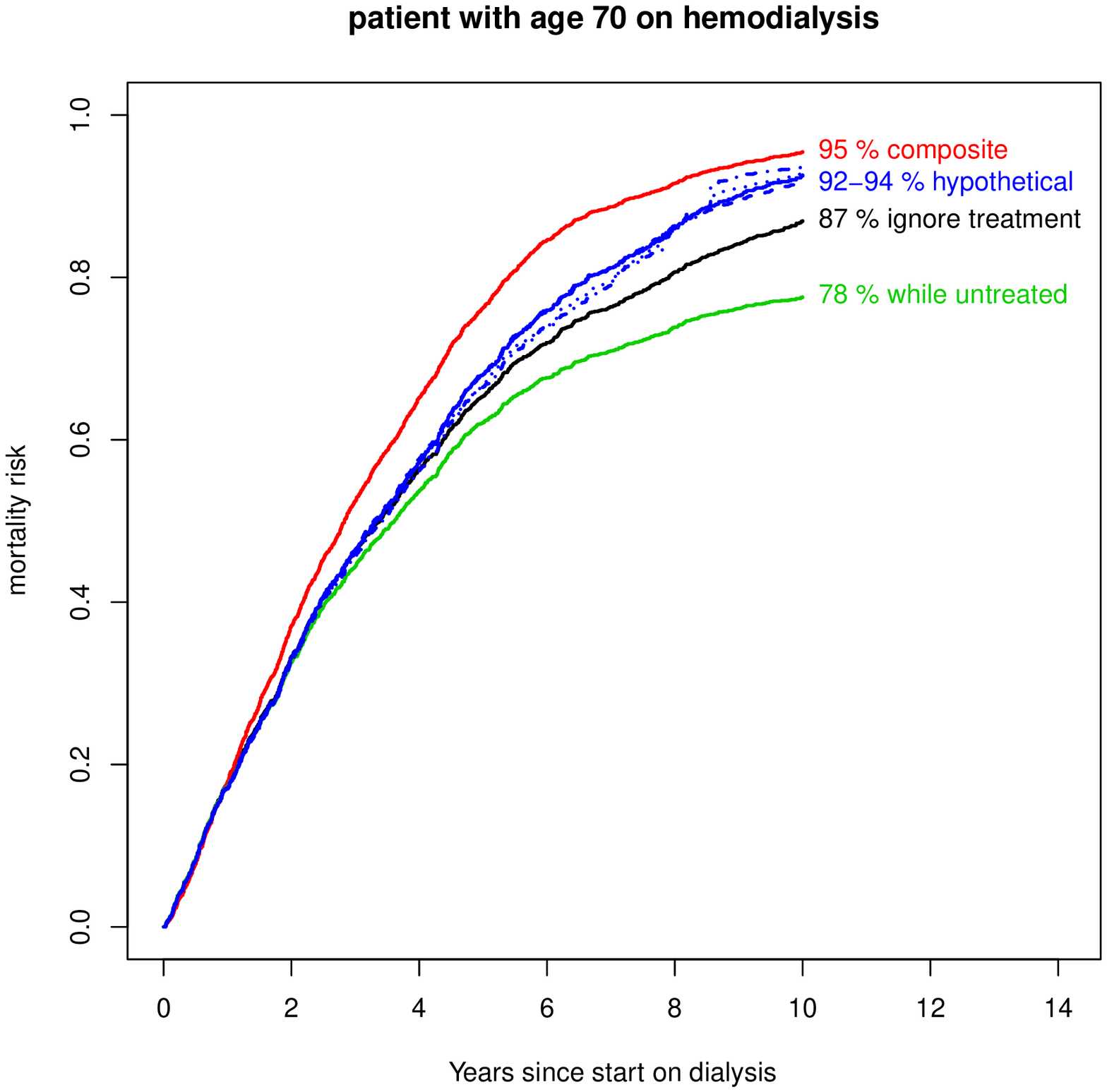}
\caption{Predicted mortality curves and 10 year mortality risks for patients aged 50 and 70 on hemodialysis. red: composite, green: while untreated/cumulative incidence, black: ignore treatment, solid blue: hypothetical - censor at treatment, dashed blue: hypothetical - modelling treatment, dotted blue: hypothetical - censor at treatment + IPW, dotdash blue: hypothetical - modelling treatment + IPW }
\label{fig:neco}
\end{figure}

In Figure \ref{fig:neco} the predicted 10 year mortality curves are presented for a patient of age 50 and for a patient of age 70, both starting on hemodialysis.  Each curve represents a different type of mortality risk. Several observations can be made from the curves. The risks obtained from the ``composite'' strategy are highest while the curves from the ``while untreated'' strategy were lowest for most of the follow up times. This is according to expectation as the ``composite'' strategy counts every transplanted patient as event, while the ``while untreated'' counts transplantation as non-event. The other two strategies infer that part of the transplanted patients will reach the event, either according to observed deaths after transplantation in the ``ignore treatment'' strategy, or according to what would be expected if these patients were not transplanted in the ``hypothetical strategy''. The fact that the ``while untreated'' strategy does not yield the lowest predictions at all times can be explained by the modelling assumptions (proportionality of covariates is acting on different scales, composite hazard versus cause specific hazard versus marginal hazard). The composite curves for 50 and 70 years-old were very similar because younger patients have a lower probability of dying but a higher probability of getting a transplantation. The four curves belonging to the ``hypothetical'' strategy are higher than those from the ``ignore treatment'' strategy, indicating that the current transplantation policy reduces mortality compared to a hypothetical scenario where nobody would receive a transplant. This is more apparent at age 50, since more patients are transplanted at that age. The curves for patients starting on peritoneal dialysis were very similar to the curves for hemodialysis and are therefore not shown.

Our focus is on the interpretation of the different prediction estimands. Below we sketch how our results could be used in a fictitious conversation between a doctor and a patient of age 50 starting on hemodialysis.

\noindent\fbox{%
    \parbox{\textwidth}{%
 \begin{itemize}
\item doctor: You have progressed to end stage renal disease, meaning your kidneys no longer function sufficiently. My advice would be to start hemodialysis.
\item patient: What is my prognosis on hemodialysis?
\item doctor: If we did not perform kidney transplantations, our best estimate is that 59 to 62\% of patients your age would die within 10 years.  (\textbf{hypothetical})
\item patient: Ok, but what about my prognosis given that I may receive a kidney transplantation?
\item doctor: With availability and allocation of transplants like in recent years, about 43\% of patients dies within 10 years. (\textbf{ignore treatment})
\item patient: So, will I get a transplant in time?
\item doctor: I cannot say, we need a matching donor and there is a waiting list. Again assuming that availability and allocation of transplants does not change, in the next 10 years, you have about 34\% chance of dying before getting a transplant. (\textbf{while untreated})
\item patient: What are the chances I will survive for 10 years and still be on dialysis?
\item doctor: With unchanged transplant availability and allocation, you have a 5\% chance to still be alive and without transplant in 10 years. (1 minus \textbf{composite})
\end{itemize}
   }%
}

We note that our simplified model with only two predictors is not meant for use in clinical practice. Also for simplicity, we omitted to report uncertainty intervals around the predictions. This example serves as an illustration that the different strategies of handling treatment start after baseline answer different risk questions.

\section{Discussion}

Starting treatment after baseline is very common in risk prediction settings. We argue that the way treatment is dealt with should not be degraded to `just a technical analysis choice'. In fact, different strategies may yield very different risk predictions. If not dealt with up-front, the choice may sneak in by the choice of analysis rather than being identified intentionally as the prediction estimand of interest. Decisions about how one wants to handle treatment initiation should be prespecified, based on which interpretation of risk is most appropriate. In some cases, multiple questions may be of interest and therefore investigators may choose to estimate more than one of the four predictimands we described here.

Being clear about the prediction question of interest in the context of post-baseline occurrences is not only important for treatment initiation. Any post-baseline behaviour or event that may be modifiable could be considered in light of our proposed framework. For example, in the transplantation setting, another modifiable post-baseline event that could be considered is patients who stop dialysis due to recovery. (In our analyses patients were censored in case of ceasing dialysis due to recovery, implying we used the ``hypothetical'' strategy with respect to this event.)

A recent systematic review on prediction models for on-dialysis mortality identified 16 models studying time to death \cite{Ramspek}. Five of these used the Cox model with censoring at transplantation, implicitly targeting the ``hypothetical'' prediction estimand. None of these five studies explained the consequences of censoring on treatment start for the interpretation of the calculated risk, and none paid attention to the non-informative censoring assumption. Three other models included death after transplantation in their outcome (``ignore treatment'' strategy). Three studies excluded patients who received transplants after baseline from their analyses, leading to predictions that are not generalizable. Five did not write anything about how they dealt with transplantation, essentially rendering risk numbers that cannot be interpreted.

Whereas causal inference research is typically strictly distinguished from prediction research \cite{Hernan2019}, we show in our paper that when predicting in the presence of modifiable events after baseline such as treatment initiation, methods from both domains are needed. A causal inference model aims to quantify what the counterfactual or potential outcomes of patients would be with and without an intervention and infers a causal effect of intervention from that. 
A prediction model aims to provide correct predictions of an outcome given a set of prognostic factors that do not have to be causally related to the outcome. The ``hypothetical'' prediction estimand could be classified as a type of counterfactual prediction, since we predict potential outcomes `if the world were different', namely, if no one receives treatment \cite{Hernan2019}. Several untestable assumptions are needed here, as described in Section \ref{sect:estimators_hyp}. The other three prediction estimands give `real world' predictions and can be estimated from a development dataset without untestable assumptions. However, when predicting for new patients using these three strategies, we assume that similar treatment assignment policies apply as in the development cohort. This is also a very strong assumption that cannot be tested upfront. The prognostic factors used in a clinical prediction model do not have to be causally related to the outcome, however the predictions they render will only be valid if the treatment policies that patients to whom the prediction model is applied are 1) clearly defined in the prediction estimand and 2) similar as in the development cohort (except for the ``hypothetical'' strategy).

Our focus in this paper is on the role of treatment in clinical prediction models. We have sketched how different strategies of handling treatment lead to different prediction estimands. The definition of the role of treatment and other intercurrent events is necessary but not sufficient to define a prediction estimand. Other aspects that need to be defined are the target population/setting (e.g., patients with end stage renal disease), the relevant outcome with an appropriate time horizon (e.g, 10-year mortality) and a time-point at which the prediction will be made (e.g., at start of dialysis) \cite{Pajouheshnia2018}.

Throughout the paper we have referred to a single treatment and assumed that treated patients remained treated throughout follow up. Usually many types of treatment are relevant for patients. For instance, data used for the development of a cardiovascular risk model may contain information on patients who start using statins, patients who start using antihypertensives or lipid lowering drugs, patients following a particular diet etc. Depending on the goal of the risk prediction, a choice should be made as to how each is handled. Typically a mixture of approaches will be used. Many treatments will be considered `care as usual' with assignment policies that are considered stable over time and can be handled as background according to the `ignore treatment' strategy. However, if for example the prediction model is aimed to input to the question of whether new patients should or should not be given statins, then a different strategy should be used for statins. For each treatment, an explicit choice should be made as to which prediction estimand is targeted and appropriate attention should be given to the necessary assumptions for estimating this. When only few patients are treated (e.g., due to a short prediction horizon) or when treatment effects are small relative to the effect of the other prognostic factors in the model, the numerical differences between the strategies will be less pronounced than in our transplantation example, but may still be relevant.

In case patients switch between `off' and `on' treatment multiple times during follow up, the definition and analysis approach of the ``ignore treatment'' and ``hypothetical'' strategy can stay unchanged. An application of the ``hypothetical'' strategy in such an `on' and `off' switching situation is presented in \cite{Pajouheshnia}. In the ``composite'' strategy it seems most sensible to count the first occurrence of a treatment episode as an event, but recurrent event approaches could also be considered \cite{Cook}. The definition of the ``while untreated'' strategy could be extended to represent the risk of the event of interest during all untreated episodes, so not only up to the first treatment episode as in the current definition.

An aspect of prediction modelling that was not addressed in our paper is assessment of predictive performance, i.e., model validation. Standard methods to validation of predictions apply to the ``ignore treatment'' and ``composite'' predictimands. For the ``while untreated'' strategy, methods suitable for competing risks analyses are needed, see for instance \cite{Saha,Schoop,Zhang}. Validating predictions generated with the ``hypothetical'' strategy is more involved since also in a validation dataset there will likely be patients who start treatment after baseline. Validation in such a setting may require similar assumptions and estimation techniques as during estimation of the hypothetical predictions \cite{Pajouheshnia2017b}. This warrants further research.

There might be a trade-off between relevance of an estimand and the assumptions that one is willing to make in order to produce it. Due to its strong and untestable assumptions some authors have argued against using a ``hypothetical'' estimand saying one can better `stick to this world' \cite{Anderson}. We argue that the future use of the prediction model should drive the predictimand choice. One should start by defining a clear estimand before considering how to compute it. When there is much uncertainty on the used assumptions, sensitivity analyses could be performed to assess the degree of uncertainty in the predictions \cite{Lash}.

In any case, when using a prediction model in clinical care, the meaning of predictions presented to patients should be unequivocally clear. Our predictimand framework can help researchers explicating what risk is targeted by their model.

\section{Funding}

Tim Morris was supported by the Medical Research Council (grant numbers MC\_UU\_12023/21 and MC\_UU\_12023/29).

\end{document}